\documentclass{PoS}

\usepackage{amsfonts, amsmath, amssymb, amsbsy}
\usepackage{epsfig}
\usepackage{graphicx}
\usepackage[numbers]{natbib}
\usepackage{fontenc}
\usepackage{times}
\usepackage{mathptmx}
\usepackage{amssymb}

\title{Results from the long-term Fermi-LAT observations of the white dwarf binary pulsar AR Scorpii}

\ShortTitle{White Dwarf Pulsars and Gamma-rays}

\author{Q. Kaplan$^{1}$, P. J. Meintjes$^{1}$,  K. K. Singh$^{1,2}$,  H. J. van Heerden$^{1}$, F. A. Ramamonjisoa$^{1}$,  \speaker{I. P. van der Westhuizen$^{1}$}\\
	1.) Department of Physics, University of the Free State, PO Box 339, Bloemfontein, 9300\\
	2.) Astrophysical Sciences Division, Bhaba Atomic Research Centre, Mumbai, India, 
	400085 \\ 
	
	E-mail: \email{KaplanQ@ufs.ac.za}}

\abstract{The discovery of the  white dwarf binary system AR Scorpii (AR Sco) with its fascinating   non-thermal dominated multi-frequency emission has sparked renewed interest in potential high energy gamma-ray emission from white dwarf pulsars. The Spectral Energy Distribution (SED) below and above optical shows evidence of non-thermal synchrotron emission, with pulsed emission in optical and X-ray bands at the white dwarf spin period ($P_{*} = 117 \, \mbox{s}$) as well as a beat period ($P_{\rm b} = 118.1 \, \mbox{s}$) with the binary period. From an energy perspective, the highly magnetic rotating white dwarf can accelerate particles to TeV energies. In this study, a search for high energy gamma-ray emission was conducted between 100 MeV - 500 GeV by analysing the newly available Fermi-LAT Pass 8 data with the new Fermi 1.0.1 Science Tools. Binned likelihood analysis was done using power law, broken power law and log parabola models. From the selected Region of Interest (ROI) centred on AR Sco's position, we calculated a significance of $\sqrt{TS}\leq$ 3.87$\sigma$  for the integrated gamma-ray activity between 100 MeV - 500 GeV at a photon flux level of 0.486 $\pm$ 0.261 x 10$^{-8}$ photons cm$^{-2}$ s$^{-1}$ using the broken power law model. This resulted in a 3$\sigma$ upper-limit detection from the position of AR Sco. The location of AR Sco inside	the Rho Ophiuchi (Rho Oph) molecular cloud complex combined with the poor spatial resolution	of Fermi-LAT,  complicates any positive identification of low-level gamma-ray	activity at the location that coincides with the position of AR Sco. }

\FullConference {36th International Cosmic Ray Conference - ICRC2019\\
	July 24th - August 1st, 2019\\
	Madison, Wisconsin, USA}

\begin{document}
	
	\section{Introduction}
	
	Since the discovery of the peculiar close binary system AR Scorpii (AR Sco) in 2016, (\cite{Marsh-2016}) which displays non-thermal dominated multi-frequency emission from radio to X-rays (see Fig. 1), it has been the focus of intensive observational and theoretical investigation. This system consists of a 0.8 - 1.29 $M_{\odot}$ highly magnetic white dwarf with a surface field $B \leq 500 \, \mbox{MG}$  (e.g. 
	\cite{Buckley-2017, Buckley-2018}) orbiting an 0.28-0.45 $M_{\odot}$ M5 dwarf star with an orbital period of $P_{\rm orb} = 3.56 \, \mbox{hours}$. Assuming the secondary star fills its Roche lobe, the distance to the system is estimated to be approximately $d \sim$ 116 pc \cite{Marsh-2016}.
	
	\begin{figure}[h]
		\begin{center}
			\includegraphics[width=0.5\textwidth]{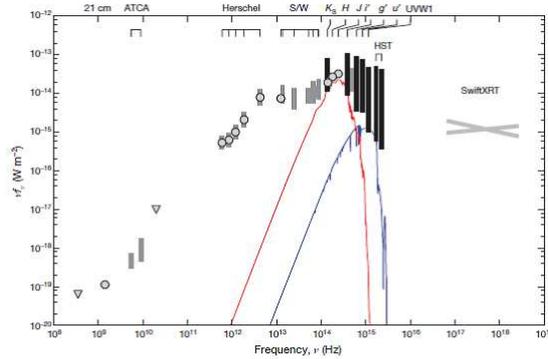}	
			
			\caption{The multi-frequency Spectral Energy Distribution (SED) of AR Scorpii from radio to X-ray frequencies.  The red and blue curves represent model atmospheres of the red M5 secondary dwarf star (T = 3100 K)  and white dwarf (T = 9750 K) respectively. Adopted from \cite{Marsh-2016}.}	
			\label{fig1}
		\end{center}
	\end{figure}

	In the optical part of the spectrum (see \cite{Marsh-2016}) the system reveals strong brightness variations every 117 s (i.e. pulsations), which is the spin period of the white dwarf, superimposed on the orbital period as well as strong pulsations at 118.2 s, which is a beat period between the spin period and the binary orbital period (see Fig. 2). Both the spin and beat periods are clearly visible on a power spectrum (see Fig. 3), especially at the first harmonic where a clear differentiation can be distinguished between the spin and beat periods  \cite{Marsh-2016}. 
	
	\begin{figure}[h!]
		\begin{center}
			\includegraphics[width=0.6\textwidth]{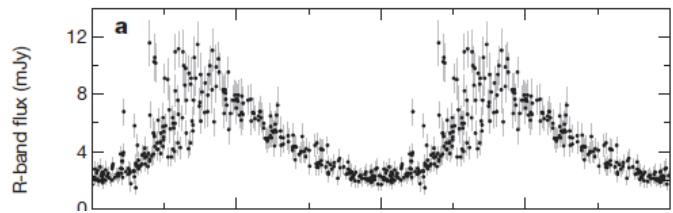}	
			
			\caption{The orbital modulation from  AR Scorpii in R-band, showing ellipsoidal variations that may be due to a Rochle lobe filling (or nearly  lobe filling) secondary star. Superimposed on this orbital light curve, optical pulsations can be seen at the spin period (117 s) and beat period (118.2 s) with the binary system. Adopted from \cite{Marsh-2016}.}	
			\label{fig2}
		\end{center}
	\end{figure}

	\begin{figure}[h]
		\begin{center}
			\includegraphics[width=0.7\textwidth]{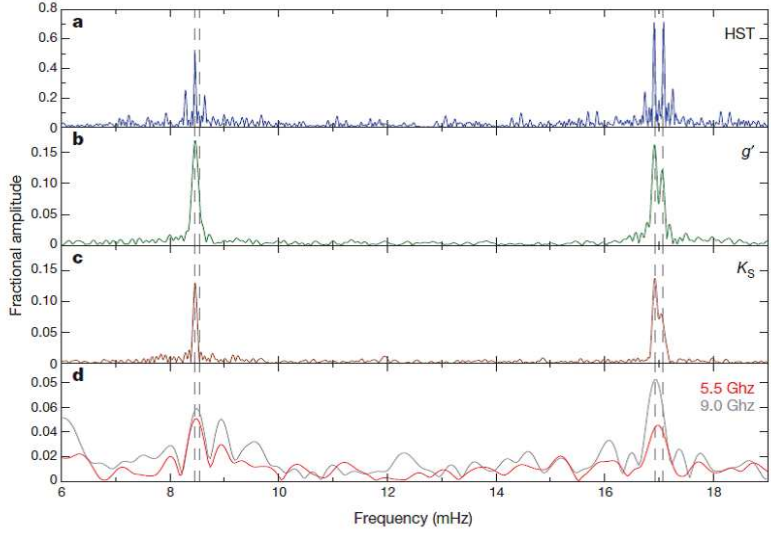}	
			
			\caption{Periodograms from radio to optical (HST), all revealing the strong double pulse beat period between the spin and orbital periods of the system. The beat pulse and the spin frequency are most prominent at the first harmonic, close to 17 mHz. The peak on the left represents the beat while the peak on the right represents the spin frequency. Adopted from \cite{Marsh-2016}.}	
			\label{fig3}
		\end{center}
	\end{figure}
	
	\newpage
	High-speed all-Stokes optical polarimetry utilizing the South African Astronomical Observatory (SAAO)  HIPPO Polarimeter \cite{Potter-2010} on the 1.9 m telescope reveals strong linear polarization at levels up to 40 $\%$ with significantly lower levels of circular polarization ($< 10 \%$), see Fig. 4, \cite{Buckley-2017, Buckley-2018}. The nature of the linear polarization profile is consistent with synchrotron emission produced in the magnetic field of a rotating magnetic dipole, strengthening the notion that the white dwarf in AR Sco is a white dwarf pulsar.   
	
	Also, the notion that nearly the entire Spectral Energy Distribution (SED) may be dominated by non-thermal emission sparked enormous interest among theorists (e.g. \cite{Geng-2016, Katz-2017}). A favorite approach was to model the emission as being the result of the interaction between a highly magnetized white dwarf pulsar and an M-type secondary dwarf star (see Fig. 5). It is proposed that the WD is a nearly perpendicular rotator where both open field lines sweep past the M-type in each rotation period. The interaction between the particle beam streaming out from the open field line regions and the M-type wind would lead to the formation of a bow shock, where the relativistic electrons can be accelerated by the large magnetic field to produce possible synchrotron radiation in the magnetic field of the magnetosphere. However, it has also been shown that very high energy gamma-ray production through a hadronic channel like $\pi^\circ$ is also possible \cite{Bednarek-2018}, as well as inverse Compton scattering.
	
	\begin{figure}[h]
		\begin{center}
			\includegraphics[width=.6\textwidth]{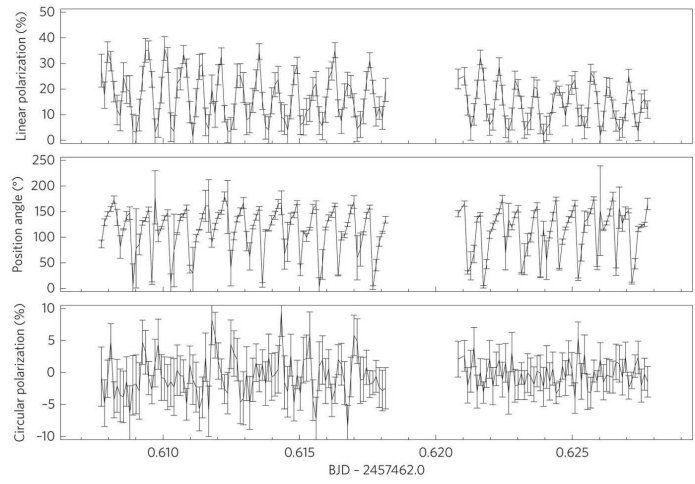}	
			
			\caption{Time series polarimetry (10s bins) that shows the degree of linear polarization, position angle of linear polarization and the degree of circular polarization. Adapted from \cite{Buckley-2017}.}	
			\label{fig4}
		\end{center}
	\end{figure}
	
	\begin{figure}[h]
		\begin{center}
			\includegraphics[width=1.0\textwidth]{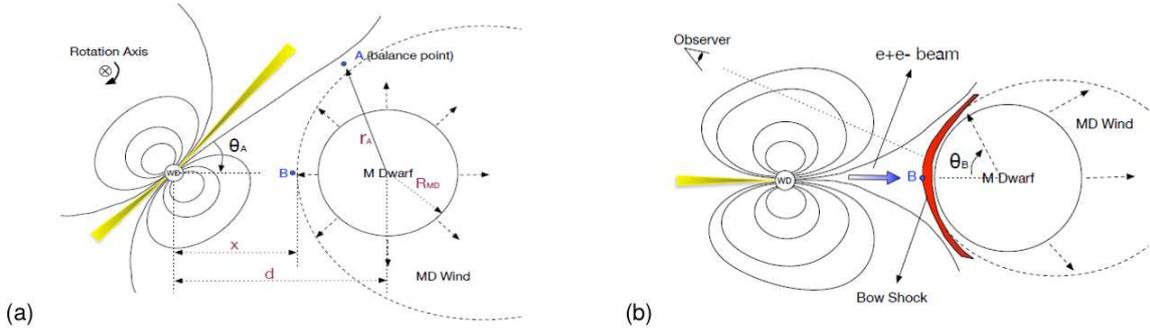}	
			
			\caption{Proposed schematic illustration of the WD/MD binary in AR Sco, where (a) shows the increase of emission from the white dwarf as the electron/positron beam approaches the M-dwarf wind (point A), and (b) shows the episode when the open	field lines of the WD interacts with the M-dwarf to form a bow shock, where electrons are accelerated to relativistic energies. Adopted from \cite{Geng-2016}.}	
			\label{fig5}
		\end{center}
	\end{figure}
	
	Recent studies of the pulsed nature of the emission in X-rays using XMM-Newton data \cite{Takata-2018} reveal pulsed emission consistent with the white dwarf spin period with a pulsed fraction of $\sim$ 14 \%. Here it is suggested that the X-ray emission is produced mainly by a thermal plasma on the secondary star heated by magnetic pumping from the white dwarf's magnetic field lines that sweeps across it during every rotation cycle. The X-ray luminosity of $L_{x} = 4 \times 10^{30} \, \mbox{erg s}^{-1}$ is then produced mainly by the thermalized electrons on the face of the secondary star,  with the non-thermal component, peaking in IR/Optical, produced by relativistic electrons accelerated in the magnetosphere of the white dwarf \cite{Takata-2018}. It is also suggested that the relativistic electrons are trapped within the white dwarf's magnetic field lines due to the magnetic mirror effect. Both the models presented by Geng et. al. \cite{Geng-2016} and Takata et. al. \cite{Takata-2018} suggest non-thermal emission  inside this region.
	
	For  AR Sco  the  ratio of the X-ray luminosity to spin-down power, i.e. $\alpha = L_{x}/L_{\dot{\nu}_{\rm s, wd}}  \sim 10^{-3}$, is similar to another cataclysmic variable AE Aquarii and place both AR Sco and AE Aquarii in the same class as spin-powered pulsars \cite{Becker-1997}. The high ratio of spin-down power  to thermal X-ray emission  provides a reservoir for particle acceleration that makes AR Sco a possible candidate for  high energy gamma-ray emission like most other spin-powered pulsars \cite{Becker-1997}. 
	
	\section{Using  10 years of  Fermi-LAT data to search for $\gamma$-ray emission}
	
	A Fermi-LAT dataset from the past decade (4Aug2008-18Mar2019) was extracted from the Fermi Science Support Center (FSSC) in the energy range of 100 MeV to 500 GeV. By using the Pass 8 data analysis pipeline, which provides a better determination of the diffuse galactic gamma-ray emission and  a significant improvement in terms of energy resolution from previous Fermi-LAT pipelines, it was possible to do a standard Binned Likelihood Analysis on the Fermi-LAT dataset. 
	
	The analysis was performed using the recently released Fermi Tools software,  where a number of energy ranges were experimented with to fit and produce likelihood statistics for AR Sco. The event files, i.e. the photon and spacecraft files that were extracted from the FSSC site, were chosen to have a Region of Interest (ROI) of 10$^\circ$ centred at ARSco (RA:$16^{h} 21^{m} 47.28^{s}$, Dec: $-22^{\circ} 53^{'} 10.39^{''}$, J2000) in order to extract the high energy photons for analysis. During reduction, only SOURCE class events (\texttt{evclass=128}) were considered, along with the relational filter (\texttt{DATA\_QUAL$>$0 and LAT\_CONFIG==1}) to select good time intervals. All the point sources in the fourth Fermi-LAT catalogue (4FGL) located within the ROI were modelled in the spectral fits, including the latest isotropic background and galactic diffuse emission. All the point sources included in the background model file are the same as reported in the 4FGL catalogue. 
	
	As AR Sco is not listed in the 4FGL catalogue  we have followed the same methodology as (\cite{Paliya-2019}), which reported the detection of a gamma-ray flare from the high redshift blazar DA 193, by manually adding the source to the model centred at its coordinates. The associated parameters of spectral models of all the sources within a 2$^\circ$ radius were chosen to be free to vary while performing the likelihood fitting. Since AR Sco is not part of the 4FGL catalogue, it was first added by parameterizing a power law spectral shape where the prefactor and photon index of the model were allowed to vary during optimization. A power law model was chosen since it is the easiest to model and is analogous with non-thermal emission. After the power law model was added, both broken power law and log parabola were also used to do a likelihood fit. The final results are summarized in Table 1.
	
	\begin{table}[h!]
		\caption{\label{ex} Spectral model parameters obtained for AR Sco using Binned Likelihood Analysis. Optimization of AR Sco was done using a power law ($\Gamma$= - (2.7 $\pm$ 0.8)), broken power law and log parabola to determine the likelihood of detection of gamma-rays from the ROI centred on AR Sco. }
		\begin{center}
			\begin{tabular}{ccccc}
				\hline
				Spectral model & IntFlux(photons/cm$^2$/s) & N$_{pred}$ & TS & Significance \\
				\hline
				Power Law  & 0.565 $\pm$ 0.311 x 10$^{-8}$ & 1092.78 & 8.53 & $\leq$ 2.92$\sigma$ \\
				Broken Power Law & 0.486 $\pm$ 0.261 x 10$^{-8}$ & 1126.57 & 15.04 & $\leq$ 3.87$\sigma$ \\
				Log Parabola & 0.251 $\pm$ 0.179 x 10$^{-8}$ & 737.39 & 13.15 & $\leq$ 3.62$\sigma$ \\
				\hline
			\end{tabular}
		\end{center}
	\end{table}
	
	Results from a previous Fermi-LAT study, shown in Table 2  (\cite{Kaplan-2018}), using the methodology of removing sources with a TS value less than 25 showed a possible gamma-ray emission from the region. However, during the analysis process one of the sources closest to AR Sco,the unidentified source \textit{4FGL J1623.7-2315} nearly $0.6^{\circ}$ away, was not taken into account during optimization and may have  contributed to emission attributed to AR Sco. These results revealed an excess at the $\sqrt{TS}\leq$ 4.89$\sigma$ level from the position of AR Sco between the energy range of 0.1 - 500GeV. Since all the sources closest to the ROI centre has to be taken into account, the results obtained from this method (see Table 2) needs to be adjusted 
	by incorporating the associated parameters of spectral models of all the sources within a 2$^\circ$ radius mentioned earlier.

	\begin{table}[h!]
		\caption{\label{ex} Summary of  model parameters from the power law fit for AR Sco  reported in \cite{Kaplan-2018} {\bf excluding} the nearby unidentified source \textit{4FGL J1623.7-2315}.  The significance levels reported by these authors (\cite{Kaplan-2018}) are therefore perhaps too optimistic due to contamination of this near-by (~0.6$^{\circ}$) source !! }
		\begin{center}
			\begin{tabular}{cccccc}
				\hline
				Energy Bin & Spectral Index & N$_{pred}$ & IntFlux(photons/cm$^2$/s)  & TS & Significance \\
				\hline
				100MeV-500 GeV & 2.70 $\pm$ 0.82 & 2041.91 &  1.076 $\pm$ 0.318 x 10$^{-8}$ & 23.93 & $\leq$ 4.89$\sigma$ \\
				100MeV-10 GeV & 2.81 $\pm$ 0.22& 1866.78 & 0.997 $\pm$ 0.317 x 10$^{-8}$ & 19.35 & $\leq$ 4.41$\sigma$ \\
				10GeV-500 GeV & 6.03 $\pm$ 1.04 & 0.00 & 5.681 $\pm$ 1.160 x 10$^{-16}$ & 0 & 0 \\
				\hline
			\end{tabular}
		\end{center}
	\end{table}
	
	
	These new updated results (see Table 1) of our present analysis in the energy range of 0.1 - 500 GeV reveals low level emission at the $\leq$3.87$\sigma$ significance level from the position of AR Sco. After the likelihood analysis has been performed, SEDs were plotted using e.g. the power law model over 9 energy bins (see Fig. 6). The  SED that was obtained using the Fermi build-in function \texttt{bdlikeSED}. The energy flux values can be considered as as 2 $\sigma$ upper limits based on a limiting TS < 4  value per bin. The distribution of TS values per energy bin (see Figure 7) also illustrates that the most significant emission is concentrated in the energy bins below 3 GeV.

	\begin{center}
		\begin{figure*}[h!]
			\includegraphics[width=35pc,height=18pc]{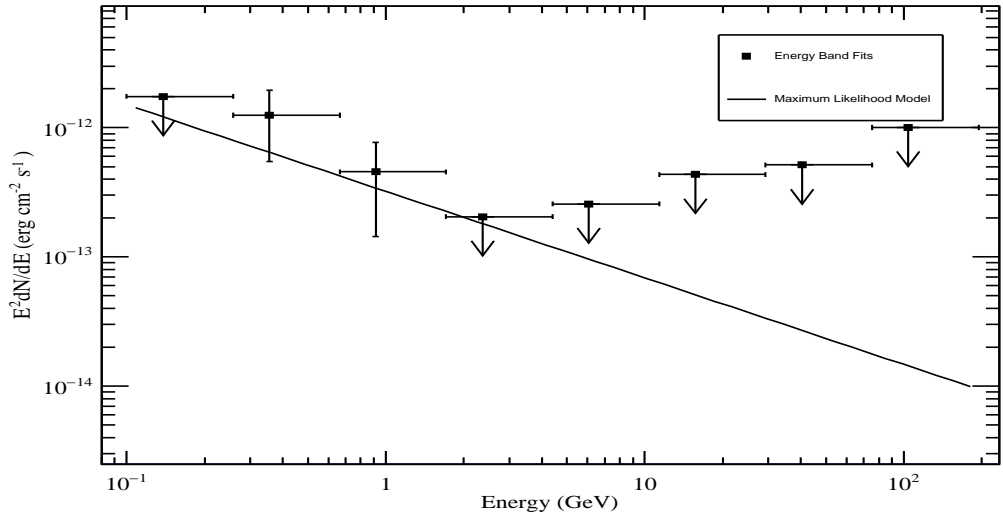}
			\caption{\label{label} Energy flux SED obtained using bdlikeSED. The flux values are 2 $\sigma$ upper limits.}
		\end{figure*}
	\end{center}
	
	\begin{center}
		\begin{figure*}[h!]
			\includegraphics[width=35pc,height=18pc]{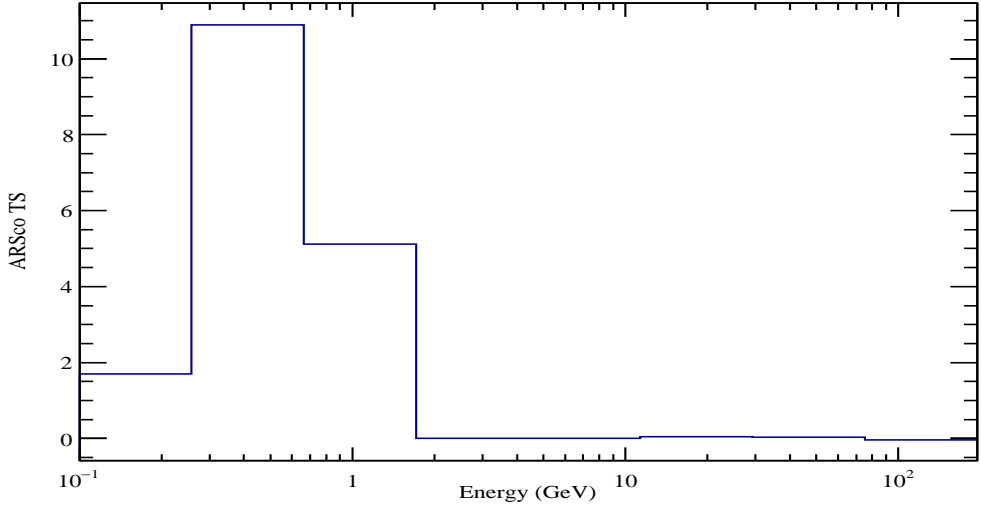}
			\caption{\label{label} TS values vs energy of $\gamma$-ray photons.}
		\end{figure*}
	\end{center}
	
	\newpage
	From the distribution of TS vs energy results, see Fig. 7, we noticed a higher TS value from the second energy bin compared to the total TS obtained, see Table 1. Likelihood analysis has been done between the 100MeV-3GeV energy range to investigate this higher TS value. The results within this energy range shows a total TS of 11.46 at a photon flux of 0.611 $\pm$ 0.482 x 10$^{-8}$ photons cm$^{-2}$ s$^{-1}$, which corresponds to the maximum TS value obtained in Fig. 7. Due to Fermi's low spatial resolution at lower energies, it may be possible for other nearby sources to still contribute to the detected emission.
	
	
	
	
	
	
	\section{Conclusions}
	
	A search for possible gamma-ray emission from AR Sco was carried out by analysing the Fermi-LAT Pass 8 dataset using the Fermi Science Tools. The results from a binned-likelihood analysis on a ROI with a radius of 10$^{\circ}$  centred on the position of AR Sco, optimized with power-law, broken power law and log parabola models,  revealed low-level gamma-ray flux between 100 MeV - 500 GeV at the 3.87$\sigma$ level by considering the broken power law as the best fit. This is below the 5$\sigma$ level heralding a possible detection. Due to the complexity of the surrounding region and the Rho Oph molecular cloud, as well as the poor spatial resolution of the Fermi-LAT telescope at lower energies, it is difficult to pinpoint the exact origin of this low level gamma-ray emission. Unbinned likelihood analysis is underway to try and maximise the probability of detection from the emitting region.      
	
	
	
	

	

\end{document}